\documentclass[aps,pra,article,twocolumn,showpacs,longbibliography,preprintnumbers,amsmath,amssymb,superscriptaddress]{revtex4-2}
\date{\today}
\usepackage{epsfig}
\usepackage{subfigure}
\usepackage{graphicx}
\usepackage{dcolumn}
\usepackage{bm}
\usepackage[colorlinks,linkcolor=blue,hyperindex,CJKbookmarks]{hyperref}
\usepackage{float}
\usepackage{hyperref}
\usepackage{comment}
\usepackage{color}

\usepackage[utf8]{inputenc}

\begin{document}

\title{Chalcogenic orbital density waves in weak and strong coupling limit}

\author{Adam K\l{}osi\'nski}
\email{adam.klosinski@fuw.edu.pl}
\affiliation{%
\mbox{Institute of Theoretical Physics, Faculty of Physics, University of Warsaw, Pasteura 5, PL-02093 Warsaw, Poland}
}%
\author{Andrzej M. Ole\'s}
\affiliation{%
\mbox{Institute of Theoretical Physics, Jagiellonian University, 
Profesora Stanis\l{}awa \L{}ojasiewicza 11, PL-30348 Krak\'ow, Poland}
}%
\affiliation{%
Max Planck Institute for Solid State Research, Heisenbergstrasse 1, D-70569 Stuttgart, Germany
}%
\author{Cliò Efthimia Agrapidis}
\affiliation{%
\mbox{Institute of Theoretical Physics, Faculty of Physics, University of Warsaw, Pasteura 5, PL-02093 Warsaw, Poland}
}%
\author{Jasper van Wezel}
\affiliation{%
Institute for Theoretical Physics Amsterdam, University of Amsterdam, Science Park904, 1098 XH Amsterdam, The Netherlands
}%
\author{Krzysztof Wohlfeld$\,$}
\affiliation{%
\mbox{Institute of Theoretical Physics, Faculty of Physics, University of Warsaw, Pasteura 5, PL-02093 Warsaw, Poland}
}%

\date{\today}
\begin{abstract}
Stimulated by recent works highlighting the indispensable role of Coulomb interactions 
in the formation of helical chains and chiral electronic order in the elemental chalcogens, we explore the $p$-orbital Hubbard model on a one-dimensional helical chain. By solving it in the Hartree approximation we find a stable ground state with a period-three orbital density wave. We establish that the precise form of the emerging order strongly depends on the Hubbard interaction strength. 
In the strong coupling limit, the Coulomb interactions support an orbital density wave that is qualitatively different from that in the weak-coupling regime. We identify the phase transition separating these two orbital ordered phases, and show that realistic values for the inter-orbital Coulomb repulsion in elemental chalcogens place them in the weak coupling phase, in agreement with observations of the order in the elemental chalcogens.
\end{abstract}
\maketitle

\section{Introduction}

\subsection{Orbital {\it versus} spin and charge density waves}

It is well-known that spin- or charge-density waves can form in 
the ground states of Hubbard models. Such density waves are 
triggered by the Coulomb repulsion, which, together with appropriate 
nesting conditions, opens a gap in the electronic band structure at 
the Fermi level and stabilizes the spin or charge density waves at 
specific fillings.
Perhaps one of the best-known examples here are the spin and charge 
density waves 
of the extended single-band one-dimensional (1D) Hubbard 
model at half-filling~\cite{Hirsch1984, vanDongen1994, Tsu2002}.
These become stable at infinitesimally weak interactions. Moreover,
their physics in the weak- and strong-coupling limits, 
although distinct in details, is qualitatively similar.

Here we investigate a distinct type of 
density wave: the orbital density wave, consisting of a periodic 
modulation of the distribution of electrons between orbitals, 
keeping the charge and spin densities constant. 
We establish that this type of orbital density wave emerges as the 
ground state of a particular Hubbard model with orbital degrees of 
freedom. We show that considering a realistic orbital Hubbard 
model yields orbital order in a way that is qualitatively 
distinct from the typical spin and charge density waves.

\subsection{Orbital density wave in the chalcogens}

In contrast to the orbital order established in Mott insulators, like the cooperative Jahn-Teller effect~\cite{goodenough}, orbital density waves are currently known to exist in only a few materials. Whereas there have been suggestions that some of the dichalcogenides, 
such as for example 1T-TiSe$_2$ or 2H-TaS$_2$, can support orbital density waves~\cite{vanWezel2011, vanWezel2012}, perhaps the simplest case concerns the two elemental chalcogens--- selenium and tellurium~\cite{fukutome1984,shimoi1992, vanWezel2010, silva2018}.
Selenium and tellurium crystals have long been known to be semiconducting and 
to consist of weakly coupled helical chains of atoms, accompanied by a 
`chiral order', at ambient pressure \cite{vonHippel1948,reitz1957,Tanaka2010,Dem20}
---though at high pressure both elements superconduct~\cite{Aka92, Str97}.

\begin{figure*}[t!]
  \vskip -0.5cm
  \includegraphics[keepaspectratio=true,width=0.45\textwidth]{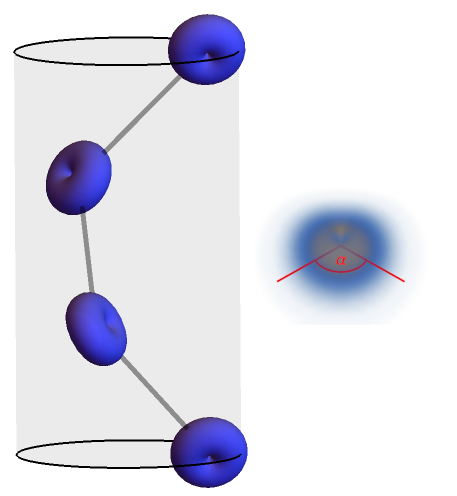}%
  \includegraphics[keepaspectratio=true,width=0.45\textwidth]{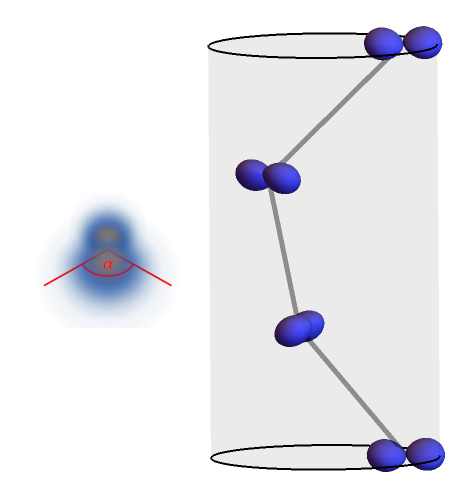}
  \vskip -.5cm
\leftline{\hskip 0.2cm {\Large (a)} \hskip 9cm {\Large (b)}}
  \vskip .3cm
\caption{
Visualisation of two possible orbital density waves with period-3
that can in principle become stable in the helical chains of elemental chalcogens:
(a) The `2-1-1' orbital density wave with two orbitals
being partially unoccupied, a configuration stable already in the non-interacting case;
(b) The `2-2-0' orbital density wave with one type of $p$ 
orbital unoccupied and supported by the inter-orbital
Coulomb repulsion. Depicted here is the probability density for the wave function 
of the {\it unoccupied orbital},
with stronger opacity indicating a higher probability density
as obtained in this paper for the orbital Hubbard model \eqref{hamiltonian} with Coulomb repulsion $U=0$
[panel (a)] and using the Hartree approximation for $U =20 t_{\sigma}$
[panel (b)], with $t_\sigma$ being the largest hopping element in the helical chain.
The bond angle in Eq.~\eqref{hamiltonian} is here taken to be $\alpha = 103 ^{\circ}$, while the hopping amplitudes obey $t_{\pi} = -t_{\sigma} /3$.
  }
  \label{charge-order}
\end{figure*}

The formation of helices naturally introduces spatially anisotropic electron hopping, which combines with the presence of an open $p^4$ valence subshell in chalcogenic atoms to give rise to the orbital density wave~\mbox{\cite{vonHippel1948, reitz1957, Matsui2014}}.
This sets the stage for explaining the formation of orbital density wave order using only a non-interacting model. This is referred to as a `valence bond' mechanism~\cite{fukutome1984}, since it originates in the lowering
of electronic kinetic energy in the 1D tight-binding model for the helical chain by a specific hybridization pattern of valence electrons between neighbouring orbitals. To be precise, on each chalcogen atom two valence electrons are assumed to reside in the two different $p$ orbitals that can hybridize with states on neighboring sites in the chain, while two other valence electrons known as the `lone pair' occupy a the remaining $p$ orbital. The pattern of orbital occupancy obtained in this simple picture is shown in~Fig.~\ref{charge-order}(a), and referred to as a `2-1-1' orbital density wave.

\subsection{The role of Coulomb repulsion}
\label{sec:introcoul}

Although the above discussion may suggest that there are no fundamental
questions related to the onset of the orbital density wave in the chalcogens, 
let us now make a `detour' and try to understand why the weakly coupled
helical chains are formed in elemental chalcogens. Note that naively
one would assume both elements to crystallise in a simple cubic structure~\cite{fukutome1984, shimoi1992, silva2018}. To resolve this issue a 
minimal 3D microscopic model~\cite{silva2018}, which builds on earlier models
\cite{fukutome1984,shimoi1992}, was recently proposed. 
It starts with the Peierls effect which triggers the formation of charge density waves with period three in the three `straight' chains formed by the $p_\alpha$ orbitals along each of the $\alpha=x,y,z$ cubic directions,
accompanied by the formation of short and long bonds in those chains.
%

Next, a {\it very small}~\cite{silva2018_2} inter-orbital Coulomb repulsion $U$ is invoked to explain the `locking' of the respective phases of each of the charge density waves. 
Then, taking into account the electronic hopping processes 
solely across the short bonds naturally leads to 
the separation of the original 3D system into quasi-1D helical chains. Altogether, this leads to no net charge modulation per chalcogen atom~\cite{fukutome1984, silva2018} and thus the charge density waves from all orbital channels combine to form orbital density waves in helical chains, see~Fig.~1 of \cite{silva2018}.

Within this picture, the helical chain thus necessarily has a nonzero inter-orbital Coulomb repulsion which stabilises an orbital density wave---however, basic calculations within a spinless model suggest (cf. 
Fig.~1 of \cite{silva2018}) that this density wave consists of a `2-2-0' pattern with
two `lone pairs' and one empty orbital on each chalcogen atom, as shown in~Fig.~\ref{charge-order}(b).

The purpose of this paper is to investigate the apparent inconsistency between
the two models discussed above. Whereas the 1D tight-binding model supports the onset of a `2-1-1' orbital density wave in each helical chain, the 3D model with finite electron-phonon and inter-orbital Coulomb repulsion explains the formation of helical chains but gives rise to a `2-2-0' orbital density wave. In particular, it is clear that if the presence of helical chains in the crystal structure of elemental chalcogens indeed relies on the presence of Coulomb interactions, these should be included in any realistic electronic model. 

This leads us to address two questions: (i)~What is the critical value of the 
Coulomb repulsion $U_{\rm crit}$ which triggers the onset of the (unrealistic) 
`2-2-0' orbital density wave in the helical chain? 
(ii) What is a realistic value of the inter-orbital Coulomb repulsion 
in the chalcogens---is it smaller than $U_{\rm crit}$ so that, despite 
the finite Coulumb repulsion, the model for the electronic band structure
of the helical chains can still support the `2-1-1' orbital density wave that is indirectly observed in the chalcogens?
Note that the answer to the above questions cannot be easily predicted 
by some kind of back-of-the-envelope calculations; for instance, in the 
well-known case of the extended 
Hubbard model~\cite{Hirsch1984, vanDongen1994, Tsu2002}, both
the charge and spin density waves are stabilised already by an infinitely small Hubbard $U$, so that \mbox{$U_{\rm crit}=0$.}

To study the role of Coulomb repulsion in the formation of the orbital density wave 
in the two elemental chalcogens, we first formulate a particular  
$p$-orbital Hubbard model on a 1D helical chain, see Sec. \ref{sec:model}. 
We then turn to the Hartree approximation, which is used to obtain 
solutions in both the weak and strong-coupling regimes, see Sec. 
\ref{sec:hfa}. In Sec. \ref{sec:TB} we present the results of the 
tight binding model which are extended by the effect of $U$ in
Sec. \ref{sec:withU}. The results are discussed in Sec. \ref{sec:dis}.
First, the orbital density wave is visualized in Sec. \ref{sec:vis}.
Next we interpret the results obtained in the weak-coupling 
(Sec. \ref{sec:odw<}) and strong-coupling (Sec. \ref{sec:odw>})
limits. We discuss the qualitative differences 
between the orbital density waves found in the different regimes of 
coupling strength in Sec. \ref{sec:dis}. The paper is 
summarized in Sec. \ref{sec:summa}, while we derive the nearest neighbor hopping matrix
in Appendix \ref{t_matrix} and verify the employed Hartree approximation 
using the preliminary density matrix renormalization group (DMRG) 
simulations in Appendix \ref{appb}.

\section{Model and Methods}
\label{sec:model}

Since the late 1940's, it has been known that the crystal structure of 
trigonal selenium and tellurium consists of loosely-coupled, 1D, helical 
chains \cite{vonHippel1948}. It can be thought of as a deformation of a 
hypothetical `parent' cubic lattice in which the bond angles are enlarged 
along helical paths through the cubic structure, as shown schematically in 
Fig.~\ref{geometry} \cite{olechna1965,chen1966,tutihasi1967,cherin1967}. 
This yields a helical chain with a period of three bonds ($\lambda=3$).
For both selenium and tellurium, the bond angle $\alpha$ has been 
experimentally determined to be $\alpha\approx 103^{\circ}$
\cite{vonHippel1948,cherin1967}. 

Both selenium and tellurium are group-16 elements (chalcogens) with the 
electron configuration $ns^2\,np^4$. Consequently, on each chalcogen 
ion in the helical chain we consider a 2/3 filled $p$-shell. To be 
able to explore both the influence of electron-electron interactions
and chain geometry on orbital order in these chains, we will construct 
a spinless three-orbital Hubbard model. We neglect the spin degree of 
freedom both to simplify the model and because spin is not expected 
to play an important role in selenium and tellurium~\cite{silva2018_2}
---see also the discussion at the end of Sec. \ref{sec:summa}.
The Hamiltonian then consists of two terms---the hopping or kinetic 
term $H_t$ and the interaction term $H_U$,
\begin{equation}
\label{hamiltonian}
  H = H_t + H_U.
\end{equation}

\subsection{The kinetic energy}

\begin{figure}[t!]
\begin{center}
\vskip -1.2cm
 \includegraphics[keepaspectratio=true,width=\columnwidth]{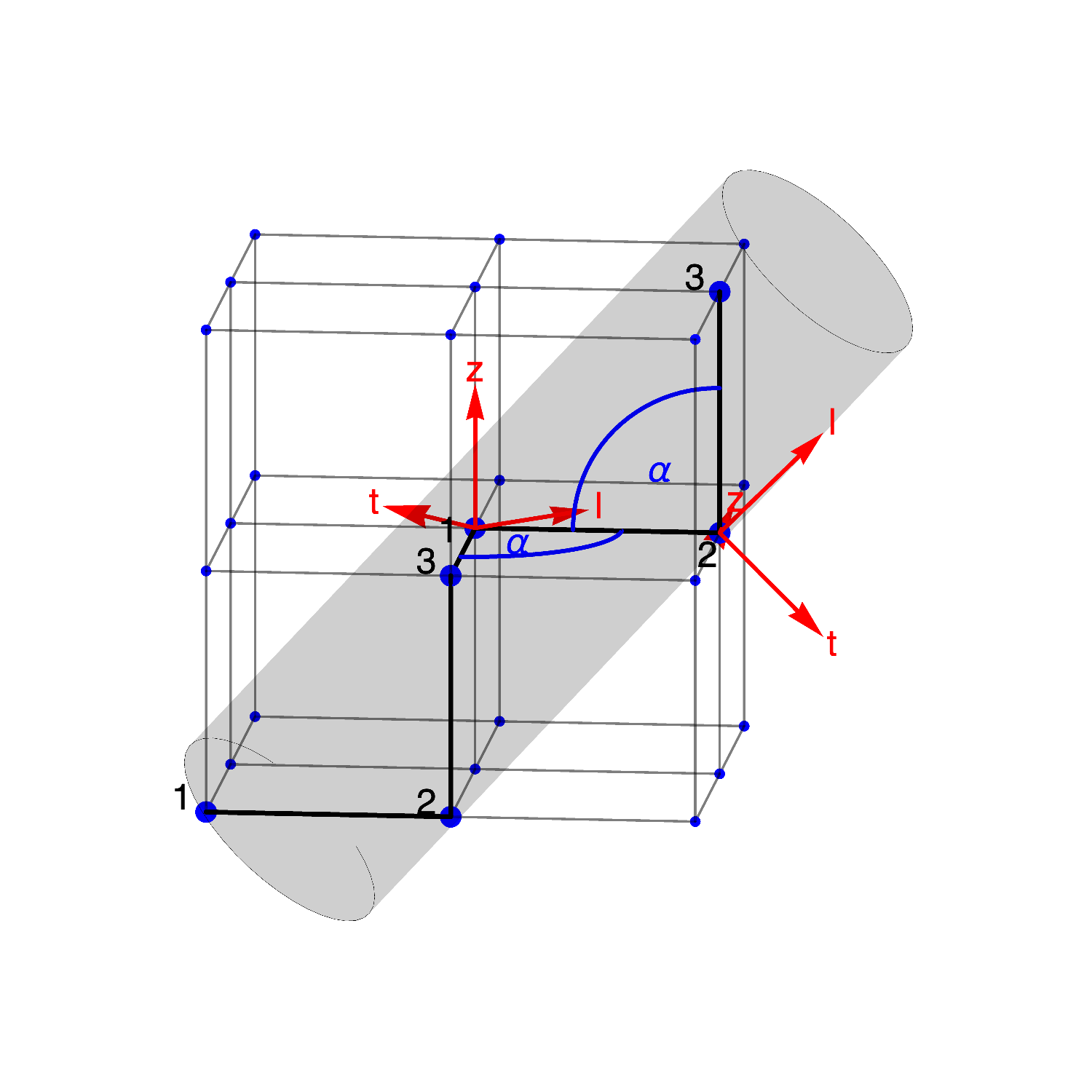}
\vskip -0.7cm
(a)
\vskip .2cm
\includegraphics[keepaspectratio=true,width=\columnwidth]{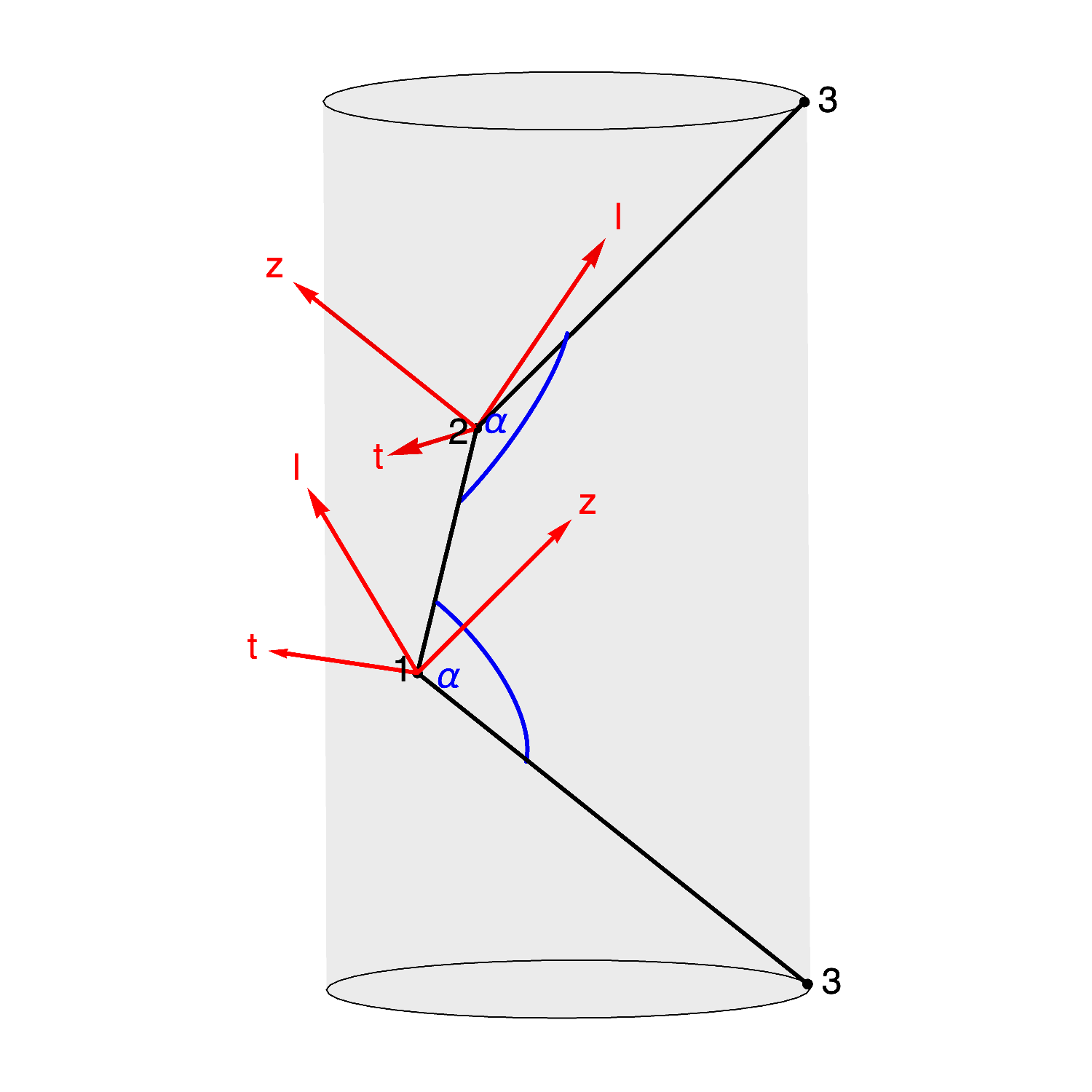}
\vskip -.0cm \hskip .5cm
(b)
\caption{
The helical chains in elemental chalcogens visualized in two possible geometries: 
(a) the idealised simple cubic case with $\alpha = 90^{\circ}$; 
(b) the realistic case with $\alpha>90^{\circ}$ 
(the selenium/tellurium structure is obtained for the bond angle $\alpha=103^{\circ}$). 
Indicated in each case are three atoms in a single ($\lambda=3$) 
period of the chain, the bond angles $\alpha$, and the local basis 
in relation to the bond angle. Looking along the chain, the three 
atoms form an equilateral triangle. The projection of the distance 
between neighboring atoms onto the chain axis depends on the bond angle.}
\label{geometry}
\end{center}
\end{figure}

Formally, we can write
\begin{equation} 
\label{hopping-term}
H_t = \sum_{i,\mu,\nu}\left(
T_{\mu,\nu}(i) c^\dag_{i,\mu} c_{i+1,\nu}^{} + H.c.\right),
\end{equation}
where $c^\dag_{i,\mu}$ ($c_{i+1,\nu}^{})$ creates (annihilates) a spinless 
electron with orbital $\mu$ $(\nu)$ on site $i$ $(i+1)$ along a helical 
chain. The orbital indices enumerate the three orthogonal $p$ orbitals 
at each site. The tunneling amplitudes between orbitals on neighboring 
sites are encoded in the hopping matrix $T_{\mu,\nu}(i)$ and depend on 
the Slater-Koster overlap integrals between the nearest neighbor $p$ orbitals~\cite{Slater1954},
\begin{equation}
t_\sigma\equiv(pp\sigma), \hskip 1cm t_\pi\equiv(pp\pi).   
\end{equation} 
Here we use $t_\sigma = 2.57$ eV \cite{olechna1965} and $t_\pi = - t_\sigma / 3$ \cite{reitz1957}. 
Note that the hopping matrix depends on the site index $i$, because  
the helical chain has three non-equivalent sites.

To derive an explicit form for the hopping matrix $T_{\mu,\nu}(i)$, 
we first need to choose an orbital basis $\{p_\mu\}$. The two most general choices 
include either picking a global basis, 
the same at each site, {\it or} considering a set of three local 
bases---one for each site in a single period of the chain. The 
Hubbard problem is much simpler if one makes the second choice, 
because the helical symmetry can then by used to render the orbital 
orientations relative to surrounding atoms the same at each site.

To this end, we choose each local basis in such a way that the lobes 
of each of the three $p_\mu\equiv\{p_x,p_y,p_z\}$ orbitals are parallel 
to the axes of a local Cartesian coordinate system. The local 
coordinates are defined by a set of three unit vectors 
$\{{\bf l},{\bf t},{\bf z}\}$ which fulfill the conditions that: 
(i)~${\bf z}={\bf l}\times{\bf t}$, 
(ii) both ${\bf l}$ and ${\bf t}$ lie in the same plane as 
the bond angle $\alpha$, 
(iii) ${\bf l}$ is perpendicular to the bisector of the bond angle 
$\alpha$ and points towards the neighboring site with highest site index, and 
(iv)~${\bf t}$ is parallel to the bisector of the bond angle $\alpha$ 
and points outwards from the bond angle $\alpha$. 
Two examples of the local coordinate systems, in relation to the bond 
angle $\alpha$, are presented in Fig.~\ref{geometry}.

This choice of local basis leads to the nearest neighbor hopping 
matrices $T_{\mu,\nu}(i)$ being the same for each site~$i$. 
Consequently, instead of working with a $(9\times9)$ hopping matrix in the 
global basis (three sites with three orbitals), we only need to 
consider a $(3\times3)$ hopping matrix in the local basis 
(one site with three orbitals). 
The trade-off in this approach is that one needs to specifically 
derive the elements of the matrix $T_{\mu,\nu}$ in terms of the 
bond angle $\alpha$ and the hopping amplitudes $\{t_\sigma,t_\pi\}$. 
Making use of the helical symmetry, this is a straightforward but 
tedious procedure, described in detail in the Appendix. 

The resulting matrix elements can be linearized with respect to the 
bond angle, around $\alpha=90^{\circ}$, which does not change the
bandwidth by more than 15\% for $\alpha$ in the range $[90^{\circ},105^{\circ}]$ 
(see Appendix).
The linearized hopping matrix is given by
\begin{align} \label{hopping-matrix}
\!& T_{\mu,\nu} =  
\frac12\! \left(
\begin{array}{ccc}
 (1+\epsilon)\,t_{\sigma }+\epsilon\,t_\pi & t_{\sigma}-\epsilon\,t_{\pi} & 
 \frac{2-\epsilon}{\sqrt{2}}\, t_{\pi}\\
-\,t_{\sigma } + \epsilon\,t_{\pi }   & 
  (-1+\epsilon)\,t_{\sigma }-\epsilon\,t_\pi  &  
  \frac{2+\epsilon}{\sqrt{2}}\, t_{\pi}
 \\
 \frac{2-\epsilon}{\sqrt{2}}\, t_{\pi} &
 -\frac{2+\epsilon}{\sqrt{2}}\, t_{\pi} & 
 -2\epsilon\,t_{\pi }  \\
\end{array}
\right)\!.\nonumber \\
\end{align}
Here $\epsilon=\alpha-\pi/2$ denotes the deviation from the simple 
cubic arrangement.

To build intuition, we first consider the special case of $\epsilon=0$, 
in the limit of $t_\pi=0$
(realistic $t_\pi$ in selenium/tellurium is expected to be around $- t_\sigma / 3$).
The non-vanishing hopping amplitudes then form a $2\times 2$ block within 
the matrix $T_{\mu,\nu}$:
\begin{equation} 
\label{hopping-matrix-simple}
 T_{\mu, \nu} (t_\pi=0, \alpha=90^{\circ}) = 
 \frac12\, t_\sigma \left(
  \begin{array}{ddd}
      1\!\! &  1 & 0 \\
     -1\!\! & -1 & 0 \\
      0\!\! &  0 & 0
  \end{array}
\!\right)\!.
\end{equation}
This result can be easily understood in terms of the simple cubic 
lattice structure. 

Since $t_\pi=0$, the only possible hopping is between orbitals of the same 
flavor aligned along the bonds. The natural basis in this case is that of 
simple cubic crystal axes. In such a (global) basis the hopping matrix is 
bond dependent. As an example, let us focus on the bond extending along 
the $\hat{x}$ axis. The hopping matrix is very simple:
\begin{equation} \label{simplehoppingmat}
 (T_x)_{\mu, \nu} = 
 t_\sigma \left(
 \begin{array}{ddd}
    1 &  0 & 0 \\
    0 &  0 & 0 \\
    0 &  0 & 0
 \end{array}
 \!\right)\!.
\end{equation}
Looking at Fig. \ref{geometry}, one can see that the matrix in Eq.~(\ref{hopping-matrix-simple}) is obtained by rotating the basis 
on the left by $45^\circ$ around the $\hat{z}$ axis, 
and the basis on the right by: (i)~$-90^\circ$ around the $\hat{x}$ axis, 
(ii) $-45^\circ$ around the $\hat{y}$ axis. It is easy to check that one gets Eq. (\ref{hopping-matrix-simple}) as a result of these transformations applied to Eq. 
(\ref{simplehoppingmat}). One can of course perform the right rotations for other 
bonds and obtain Eq. (\ref{hopping-matrix-simple}) in each case.

\subsection{The on-site Hubbard interaction}
\label{sec:U}

Since the change from the global to the local coordinate basis is 
just a local rotation, it does not affect the on-site interaction 
terms. The local Hubbard-$U$ repulsion between spinless electrons 
in different $p$ orbitals at the same site is thus written as
\begin{equation} \label{interaction-term}
  H_U = \,U\sum_{\scriptsize\begin{array}{c}
                      \mu > \nu\\
                      \mu,\nu = l,t,z
                    \end{array}}
                  \!\sum_{i}\, n_{i,\mu} n_{i,\nu}.
\end{equation}
Here we defined the number electron number operator, $n_{i,\mu}=c^\dag_{i,\mu}c_{i,\mu}^{}$ 
and
took into account 
the well-known fact that the electron-electron coupling constant 
\mbox{$U_{\mu,\nu}=U$} is the same for each pair of $p$ orbitals, cf.~Refs~\cite{Oles1983, Zhang2014}.

While in what follows we will treat the Hubbard $U$ as a model parameter 
and vary it, let us also estimate the realistic value of the inter-orbital
Coulomb repulsion between two spinless electrons in the elemental chalcogens:
\begin{align}\label{eq:uest}
U_{\rm real} &\simeq \frac14 \Big[ F^{(0)} + \frac{1}{25} F^{(2)}  \Big]
+ \frac34 \Big[ F^{(0)} - \frac{1}{5} F^{(2)}  \Big] \nonumber \\
&= F^{(0)} - \frac{14}{100} F^{(2)}  \nonumber \\
&=  U_{\rm Te} - \frac{7}{10} J_{\rm Te} \nonumber \\ 
&\approx 0.61\ {\rm eV}.
\end{align}
Here we assumed the following:
(i) in the first line of Eq.~(\ref{eq:uest}) we approximated the effective inter-orbital repulsion between two spinless electrons by a repulsion between 
two spinful electrons either in an inter-orbital singlet (3 out of 12 possible 
`inter-orbital multiplets') with energy $F^{(0)} + \frac{1}{25} F(2)$ 
or a triplet (9 out of 12 possible `inter-orbital multiplets')
with energy \mbox{$F^{(0)} - \frac{1}{5} F(2)$ ($F^{(k)}$} 
are the Slater integrals defined in a standard way, 
cf.~\cite{Zhang2014}, and the calculations of atomic multiplets are for example
available at \url{https://www.cond-mat.de/sims/multiplet/});
(ii) in the third line of Eq.~(\ref{eq:uest}) we introduced the values of 
the Coulomb repulsion parameters as estimated by Deng {\it et al.} for 
{\it solid} tellurium~\cite{Deng2007, Deng2006}:
$F^{(0)} = U_{\rm Te} \approx 1.10$ eV and $F^{(2)} / 5 = J_{\rm Te} \approx 0.7$ eV.
Note that in this way we obtain a realistic ratio $U / t_{\sigma}\approx 0.24$ 
in tellurium and that this also constitutes the lower bound for that 
ratio for selenium---since the selenium $4p$ orbitals are effectively 
`smaller' than the tellurium $5p$ orbitals and hence the value of the 
Slater integral $F^{(0)}$ should be larger in the former case.

\subsection{The Hartree approximation}
\label{sec:hfa}

To solve the Hubbard model for the helical chain, we employ the 
Hartree approximation. This means that the interaction term becomes
\begin{align} 
\label{h-f-interaction}
    H_U = U\! \sum_{\scriptsize \begin{array}{c}
                      \mu > \nu\\
                      \mu,\nu = l,t,z
                    \end{array}}\!
                  \sum_{i}& \: \Big( \langle n_{i,\mu}\rangle\: n_{i,\nu} + n_{i,\mu} \: \langle n_{i,\nu} \rangle \notag \\
                  &- \langle n_{i,\mu}\rangle \langle n_{i,\nu}\rangle\Big).
\end{align}
Because the outer $p$-shell of chalcogen atoms is 2/3-filled, we expect 
to find two spinless electrons per site. The mean fields thus need to fulfill 
the following condition at every site $i$:
\begin{align} 
\label{mfcondition}
\sum_{\mu}\;\; \langle n_{i,\mu} \rangle\, =2.
\end{align}

To solve the mean field model consisting of Eqs.~\eqref{hopping-term} 
and~\eqref{h-f-interaction}, we use the Ansatz that the ground 
state expectation values $\{\langle n_{i,\mu}\rangle\}$ 
have unbroken translational symmetry in the {\it local} basis.
That is, we look for ground states within the subspace of 
translationally invariant eigenstates that obey:
\begin{equation}
  \begin{array}{lr}
    \langle n_{i,\mu} \rangle \equiv \bar{n}_{\mu} & \text{for} \; \mu=l,t,z.\\
  \end{array}
\end{equation}
This assumption is equivalent to only considering types of order 
that respect the helical symmetry of the chain in the {\it global} 
coordinate basis. Consequently, the periodicity of the orbital 
density waves that we are looking for is encoded in our choice 
of local basis and the only unknown we need to solve for 
in the mean field analysis is the orbital occupation (again, in the local 
basis). As already discussed in the Introduction, physically this 
means that the presence of helical chains in the atomic structure 
of elemental chalcogens hardwires a preferred periodicity for any 
density wave instability. It does not, however, determine the 
amplitude and the form of any orbital density wave. That is, 
the choice of occupied orbitals resulting from the competition 
between the `2-1-1' and `2-2-0' density waves (corresponding to 
'1-0.5-0.5' and '1-1-0' ordering of spinless electrons) is still 
to be determined. These are influenced by the trigonal 
distortions (deviation of the bond angle $\alpha$ from 90$^\circ$) 
and Coulomb repulsion (represented by the Hubbard $U$).

The orbital occupation numbers $\{\bar{n}_{\mu}\}$ can be solved for in 
a self-consistent manner. Namely, we look for the lowest energy 
fixed point of the recursion relations:
\begin{widetext}
\begin{align} \label{h-f-procedure}
\left( \bar{n}_{l} \right)_{k} &= 
\frac{1}{N} \sum_q\; \left<\Phi_0(\left( 
\bar{n}_{l}\right)_{k-1}, \left(\bar{n}_t\right)_{k-1},U) \right| 
n_{q,l} \left| \Phi_0\left((\bar{n}_{l})_{k-1},
\left(\bar{n}_t\right)_{k-1},U\right)\right>,\notag \\
\left(\bar{n}_t\right)_{k} &= 
\frac{1}{N} \sum_q\; \left< \Phi_0(\left( 
\bar{n}_{l}\right)_{k-1},\left(\bar{n}_t\right)_{k-1},U) \right| 
n_{q,t} \left| \Phi_0\left((\bar{n}_{l})_{k-1},
\left(\bar{n}_t\right)_{k-1},U\right)\right>, \notag \\  
(\bar{n}_z)_k &= 2- (\bar{n}_l)_k - (\bar{n}_t)_k.
\end{align}
\end{widetext}
Here, the state 
$\left|\Phi_0\left((\bar{n}_{l})_{k-1},(\bar{n}_{t})_{k-1},U\right)\right>$ 
used in calculating the mean field values in step $k$, is the ground 
state of the Hamiltonian defined by Eqs.~\eqref{hopping-term} and 
\eqref{h-f-interaction} with mean field values 
$\langle n_{i,l}\rangle=\left(\bar{n}_{l}\right)_{k-1}$ and 
$\langle n_{i,t}\rangle=\left(\bar{n}_{t}\right)_{k-1}$, 
calculated in the step $k-1$ of the recursive procedure. 
We also used the fact that for a site-independent orbital 
occupation (in the local basis), we have
\begin{align}
\left<\Phi_0\right| n_{i,\mu} \left|\Phi_0\right> &= 
\frac{1}{N}\sum_i\left<\Phi_0\right|n_{i,\mu}\left|\Phi_0\right> \nonumber \\ 
&=\frac{1}{N}\sum_q\left<\Phi_0\right|n_{q,\mu}\left|\Phi_0\right>.
\end{align}
Note that the last equality in Eq.~\eqref{h-f-procedure} directly 
follows from Eq.~\eqref{mfcondition} and that this condition is also 
implicitly used in the first two equations in Eq. (\ref{h-f-procedure}). The calculations are performed for a 100-site chain.

\section{Results}
\label{sec:res}

\subsection{The tight binding model ($U=0$)}
\label{sec:TB}

\begin{figure}[t!]
   \flushleft{(a)}\\
  \includegraphics[keepaspectratio=true,width=0.48\textwidth]{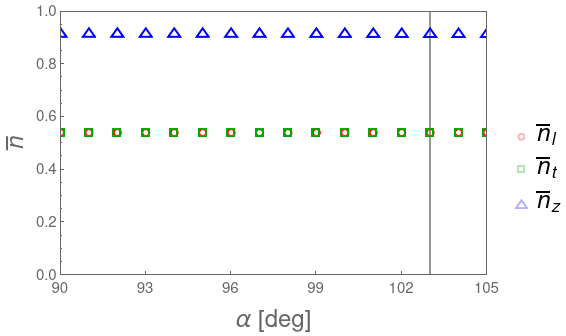}\\
  \flushleft{(b)}\\
  \includegraphics[keepaspectratio=true,width=0.48\textwidth]{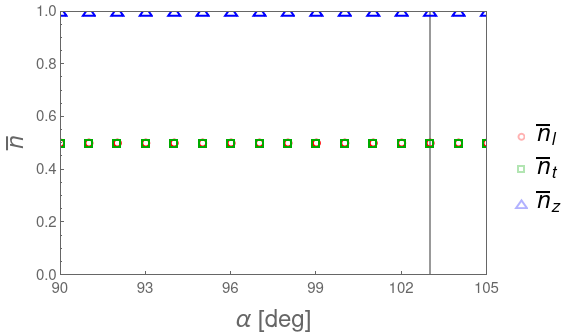}
\caption{The ground state orbital occupation $\bar{n}_\mu$ in the local 
basis with $U=0$, i.e., considering only the hopping term of 
Eq.~\eqref{hopping-term}, as a function of bond angle $\alpha$ for: 
(a) $t_{\pi}=-t_{\sigma}/3$ (for a realistic value in selenium / tellurium, 
see text) and (b)~$t_{\pi}=0$. The selenium/tellurium bond angle 
$\alpha=103^{\circ}$ is marked with a vertical line.}
  \label{non-interacting-results}
\end{figure}

In the non-interacting model with $U=0$, i.e., considering only the 
hopping term of Eq.~\eqref{hopping-term}, the orbital occupations in the 
ground state $\{ \bar{n}_\mu \}$ can be calculated exactly. 
In Fig.~\ref{non-interacting-results}(a) we show that 
the $\{ \bar{n}_\mu \}$ remain approximately constant for angles in the range 
$\alpha\in(90^\circ,105^\circ)$. 
The values at the selenium/tellurium bond angle are 
$\bar{n}_l\approx\bar{n}_t\approx 0.54$, and $\bar{n}_z\approx 0.92$. 
In Fig. \ref{non-interacting-results}(b) we show $\{ \hat{n}_\mu \}$ 
as a function of bond angle $\alpha$ when $t_\pi = 0$. As in the previous
case, the occupation numbers do not change within the pictured bond angle 
range and are very simmilar to those obtained for realistic values of $t_\pi$. 
 
Note that since both Figs.~\ref{non-interacting-results}(a) and (b) have 
been calculated using the linearized hopping matrix, results 
obtained for bond angles higher than $\approx 105^{\circ}$ differ 
quantitatively from those obtained using the full hopping matrix. 
The experimentally established bond angles in selenium and tellurium, 
however, lie well within the region where the linear approximation is valid 
(as discussed in the Appendix).

The results obtained in the non-interacting limit fully agree with the 
'valence bond picture' (as described in 
Refs.~\cite{vonHippel1948, reitz1957, Matsui2014}), which translates to 
(1,0.5,0.5) orbital occupancies in the spinless electron language.
According to this mechanism, every chalcogen atom lends a single electron 
to each of two covalent bonds, while the other two electrons remain 
in the $p_z$ orbital, normal to the bond angle plane. In the spinless
electron picture, this translates to one spinless electron distributed 
evenly between the two orbitals in the bond angle plane, $p_l$ and $p_t$, 
while the remaining electron occupies the $p_z$ orbital, 
as shown in Figs.~\ref{non-interacting-results}(a) and 
\ref{non-interacting-results}(b). 
As presented in Sec.~\ref{sec:vis} and shown in Fig.~\ref{charge-order}, 
this leads to the orbital density wave of the `2-1-1' character.

\subsection{Including the interaction term}
\label{sec:withU}

\begin{figure*}[t!]
  \flushleft{(a)} \hfill (b)\\
  \includegraphics[keepaspectratio=true,width=0.495\textwidth]{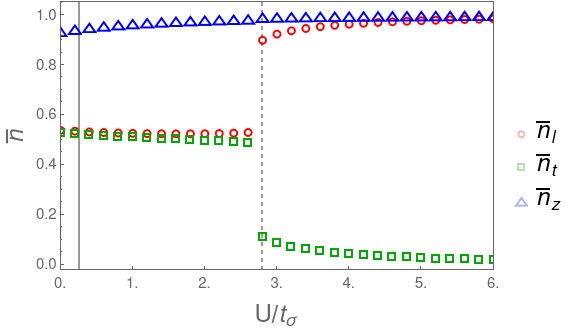}%
  \includegraphics[keepaspectratio=true,width=0.495\textwidth]{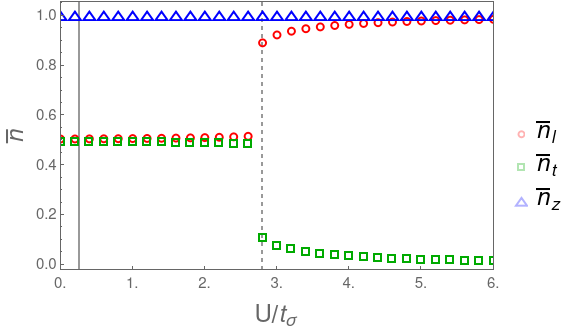}\\
  \flushleft{(c)} \hfill (d)\\
  \hskip .1cm
  \includegraphics[keepaspectratio=true,width=0.44\textwidth]{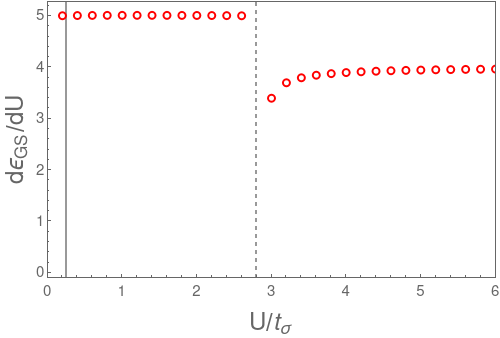}~~~%
  \hskip .7cm
  \includegraphics[keepaspectratio=true,width=0.44\textwidth]{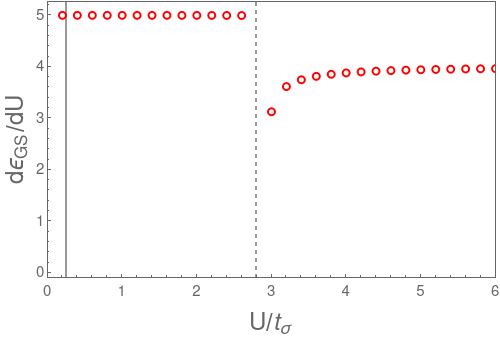}
\caption{
Evolution of the ground state properties in the Hartree approximation as 
a function of the on-site Coulomb repulsion $U$ and with hopping parameters 
$t_{\pi}=- t_{\sigma} /3$ [panels (a), (c)] or $t_{\pi}=0$ [panels (b), (d)]. 
The bond angle $\alpha$ coincides with the experimentally observed value 
of $103^{\circ}$. The top panels show the ground state orbital occupations
$\{\bar{n}_\mu\}$ in the local basis and the bottom panels show the 
derivative of the ground state energy for increasing Coulomb repulsion $U$. 
A phase transition (marked by a dashed vertical grey line) occurs at 
$U_{\rm crit}\approx 2.8t_\sigma$ for both $t_{\pi}=-t_{\sigma}/3$ 
and $t_{\pi} = 0$, with the order shifting from covalent bond formation 
(fully occupied $p_z$ orbital, the other spinless electron split evenly 
between the two orbitals in the bond angle plane) 
to localized electrons (one, $p_t$ orbital, unoccupied).
Note that  $t_{\pi}=- t_{\sigma} /3$  [panels (a), (c)]
and $U_{\rm real} \simeq 0.24 t_\sigma$ (solid grey vertical line)
are the realistic values of model parameters for the two elemental 
chalcogens, see text.
 }
  \label{interacting-results}
\end{figure*}

To study the properties of the full Hubbard model, including a 
non-zero Coulomb repulsion $U$, we employ the Hartree approximation 
described by Eq.~\eqref{h-f-interaction}. The resulting evolutions of 
the orbital occupation numbers $\{\bar{n}_\mu\}$, as well as the derivative 
of the ground state energy with respect to the Coulomb repulsion strength, 
$d\epsilon_0/dU$, are shown in Fig.~\ref{interacting-results}.
Increasing interaction strength leads to a phase transition, which 
occurs at $ U_{\rm crit}\approx 2.8 t_\sigma$ for both $t_{\pi}=- t_{\sigma} /3$ 
and $t_{\pi} = 0$. It is signalled by discontinuities in 
$d\epsilon_{GS}/dU$ [see Fig. \ref{interacting-results}(c) and 
\ref{interacting-results}(d)] and orbital occupations $\{ \bar{n}_\mu \}$ 
[see Fig.~\ref{interacting-results}(a) and \ref{interacting-results}(b)]. 

For weak interactions ($U<U_{\rm crit}$), the obtained orbital density 
wave agrees with the one discussed in Sec.~\ref{sec:TB} immediately 
above---see Figs.~\ref{non-interacting-results}(a-b).
Thus, the `valence bond picture' is valid here and the `2-1-1' character 
of the density wave is observed, see Sec.~\ref{sec:vis} 
and Fig.~\ref{charge-order}.

As the system approaches the critical value of the Coulomb 
repulsion $U_{\rm crit}$, the following effects are observed:

(i) {\bf in the bond-angle plane} the hole occupation slightly polarizes, 
favoring the $p_l$ orbital, for both zero and non-zero $t_\pi$

(ii) {\bf perpendicular to the bond-angle plane} the $p_z$ orbital 
is always occupied for $t_\pi = 0$ 
[see Fig. \ref{interacting-results}(b)], while for nonzero $t_\pi$ 
its occupation is slowly increased with increasing interaction, 
without a visible discontinuity at the transition.

For interaction strengths slightly above $U_{\rm crit}$, 
the system is approaching a saturated state, with a least-occupied 
$p_t$ orbital character. The occupation numbers change slowly upon 
further increasing $U$, so that in the infinite $U$ limit the spinless 
electrons are completely localized on the $p_t$ and $p_z$ orbitals. 
This gives (1,1,0) orbital occupation in the spinless electron model, 
which, as presented in Sec.~\ref{sec:vis} and Fig.~\ref{charge-order}, 
characterizes the orbital density wave of `2-2-0' character.

\section{Discussion}
\label{sec:dis}

\subsection{Visualising the orbital density waves}
\label{sec:vis}
 
Having found the ground state of the mean field model in the local basis, 
we can translate it back to the global basis. This allows us to clearly 
present the real-space orbital densities in selenium/tellurium chains, 
as shown already in Fig.~\ref{charge-order} of the Introduction,
for both $U<U_{\rm crit}$ and 
$U>U_{\rm crit}$. The phases in these regimes differ significantly 
in the way electronic charge is distributed over the orbitals. 

The orbital density wave stabilized in the non-interacting case and for 
all values of $U<U_{\rm crit}$, has {\it two} ($p_l$ and $p_t$) partially 
occupied orbitals lying in the bond angle plane, i.e. the orbital density 
wave has `2-1-1' character (one orbital fully occupied and two partially occupied),
and is qualitatively similar to the one depicted in~Fig.~\ref{charge-order}(a).
Interestingly, we observe that the resulting charge density is flattened in the 
direction normal to the bond angle (the $z$ direction in the local basis).
Above the critical interaction strength $U>U_{\rm crit}$ the system enters 
a different phase and the unoccupied orbital is purely of $p_t$ character 
as shown for $U=20t_{\sigma}$ in~Fig.~\ref{charge-order}(b).
In the `spinful language' this density wave corresponds to the so-called 
`2-2-0' orbital density wave (two orbitals fully occupied and one empty).

\subsection{Orbital density wave for $U<U_{\rm crit}$}
\label{sec:odw<}

\begin{figure}[t!]
  \flushleft{(a)}\\
  \includegraphics[keepaspectratio=true,width=\columnwidth]{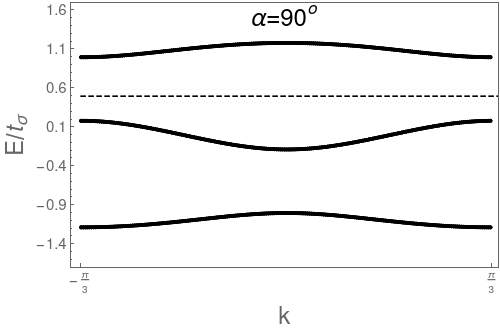}\\
  \flushleft{(b)}\\
  \includegraphics[keepaspectratio=true,width=\columnwidth]{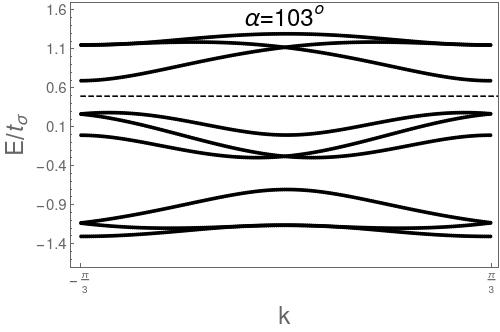}\\
\caption{The band structure for holes in the non-interacting model 
($U=0$), with hopping $t_{\pi}=-t_{\sigma} / 3$ (a realistic
value for both elemental chalcogens, see text). 
Two values of the bond angle $\alpha$ are shown: 
(a) the simple cubic case with $\alpha = 90^{\circ}$, and 
(b) the selenium/tellurium bond angle $\alpha = 103^{\circ}$. 
Dashed lines denote the Fermi energy. The band degeneracy is 
lifted as the bond angle $\alpha$ departs from $90^{\circ}$.}
  \label{non-interacting-bands}
\end{figure}

To understand the presence of an orbital density wave for 
$U<U_{\rm crit}$, it suffices to consider the exactly solvable 
non-interacting case. The evolution of the orbital occupation 
with bond angle can be then understood entirely in terms of the 
evolution of the band structure, which is shown in
Fig.~\ref{non-interacting-bands}.

In the simple cubic case, obtained for the bond angle 
$\alpha=90^{\circ}$, we see three well-separated, three-fold 
degenerate bands [see Fig. \ref{non-interacting-bands}(a)]. 
The middle three degenerate bands can be identified as having
mostly $p_z$ orbital character, while 
the other two three-fold degenerate bands are formed by linear 
combinations of the $p_l$ and $p_t$ orbitals lying in the bond 
angle plane. The degeneracy of the bands is a consequence of 
the $\alpha = 90^{\circ}$ bond angle, for which there exists a 
global orbital basis in which there is absolutely no orbital mixing, 
even with nonzero $t_\pi$. This is the basis associated with 
the three cubic crystal axes.

For bond angles $\alpha>90^{\circ}$, the degeneracy is lifted and 
for the selenium/tellurium bond angle $\alpha = 103^{\circ}$ nine 
distinct bands can be seen [see Fig. \ref{non-interacting-bands}(b)].
Nevertheless, they are well separated into three classes of bands.
The insulating character remains for the present filling of 2/3 
but the gap is somewhat reduced.

\subsection{Orbital density wave for $U>U_{\rm crit}$}
\label{sec:odw>}

\begin{figure*}[t!]
  \flushleft{(a) \hfill (b) \hfill .}\\
  \includegraphics[keepaspectratio=true,width=0.5\textwidth]{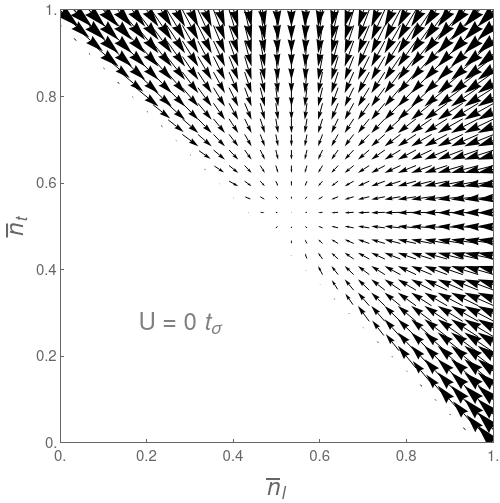}%
  \includegraphics[keepaspectratio=true,width=0.5\textwidth]{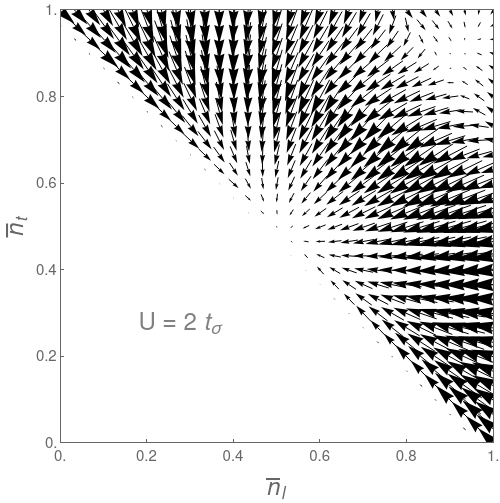}\\
  \flushleft{(c) \hfill (d) \hfill .}\\
  \includegraphics[keepaspectratio=true,width=0.5\textwidth]{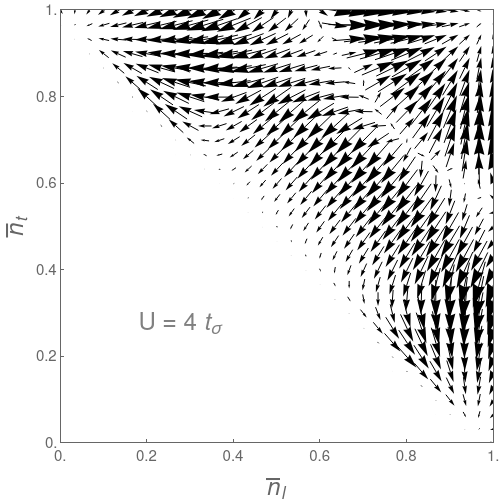}%
  \includegraphics[keepaspectratio=true,width=0.5\textwidth]{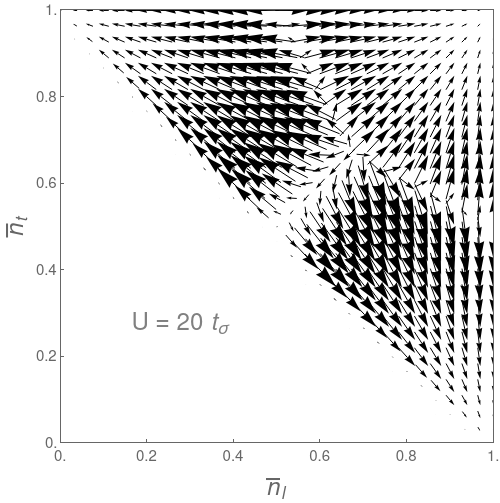}\\
\caption{The flow of the iterative procedure defined by 
Eq.~\eqref{h-f-procedure} for four different values of $U$, presented 
in terms of the vector field $\left<\Phi_0(\bar{\vec{n}})\right|\vec{n} 
\left|\Phi_0(\bar{\vec{n}})\right>-\bar{\vec{n}}$ where $\vec{n}=(n_l,n_t)$. 
The bond angle is taken to be 
$\alpha = 103^{\circ}$, while the hopping 
$t_{\pi}=-t_{\sigma} / 3$ is a realistic
value for both elemental chalcogens, see text. }
\label{fixed-points}
\end{figure*}

To understand the orbital density wave in the strong-coupling limit, 
with $U>U_{\rm crit}$, we focus on the infinite $U$ limit. Within the 
Hartree approximation, a phase transition occurs when one of two things 
happens in the flow of the iterative procedure defined by 
Eq.~\eqref{h-f-procedure}: (i)~either a new fixed point appears, 
which is also a new global energy minimum, or (ii) 
the energy hierarchy of existing fixed points changes, 
thus switching the ground state. To investigate which case is realized 
here, Fig. \ref{fixed-points} depicts the flow diagrams and fixed 
points for four values of the interaction strength.

First, in Fig.~\ref{fixed-points}(a), the non-interacting model is seen 
to have a single fixed point. As $U$ increases, the fixed point moves 
towards the $\bar{n}_t=1-\bar{n}_l$ axis [see Fig.~\ref{fixed-points}(b)]. 
This is visible as a continuous increase of $\bar{n}_z$ as $U$ departs from 
zero in Figs.~\ref{interacting-results}(a) and \ref{interacting-results}(c). 
Eventually, for $U=U_{\rm crit}$, the single fixed point vanishes, while 
three new ones emerge near the corners of parameter space, as shown in 
Fig.~\ref{fixed-points}(c). The new ground state is in the corner 
$\bar{n}_t\approx 0$ and $\bar{n}_l\approx 1$. This is the saturated 
phase with an unoccupied $p_t$ orbital.

For even larger values of $U$, the three fixed points move further 
towards the corners of parameter space [Fig. \ref{fixed-points}(d)], 
where they settle and become degenerate in the $U\rightarrow +\infty$ 
limit. Incidentally, these corner states constitute the three-fold 
degenerate ground state of the inter-orbital Coulomb repulsion on a single
chalcogen ion in the $\{p_l,p_t,p_z\}$ basis.

In the full 1D Hubbard model ($\ref{hamiltonian}$) the three-fold 
degeneracy of the ground state in the infinite $U$ limit is broken by the 
kinetic term (\ref{hopping-term}), whcih leads to the selection of $p_t$ 
as the single least-occupied orbital. This means that whereas the Coulomb 
repulsion triggers the onset of the 
orbital density wave in the strong coupling limit, the kinetic term decides 
on the precise nature of the orbital density wave.

\section{Conclusions}
\label{sec:summa}

\subsection{Summary of main results}
In this paper we studied the instabilities towards orbital density wave 
order in a $p$-orbital Hubbard model for a helical 
chain. This is the relevant geometry for the trigonal phases of the two 
elemental chalcogens selenium and tellurium 
\cite{vonHippel1948,reitz1957,cherin1967,olechna1965,Tanaka2010}. 
By considering the orbital Hubbard model for such a helical chain in 
the Hartree approximation, we showed that an orbital density wave with 
the same period as the atomic helix is stabilised, irrespective of the 
strength of the inter-orbital Coulomb repulsion $U$. The precise form of the 
orbital density wave, however, is strongly sensitive to the interaction 
strength $U$. For realistic values of both the bond angle in the helical 
chain and the ratio of the hopping amplitudes $t_\pi$ and $t_\sigma$, 
we observe a phase transition between qualitatively different orbital 
density waves at $U_{\rm crit} \approx 2.8 t_{\sigma}$.

As the main result of this work, we have shown that in the considered 
model the value of $U_{\rm crit}$ is not only nonzero, but also relatively 
large---and that the estimated value of $U_{\rm real}\approx 0.24t_{\sigma}$ 
in the two elemental chalcogens clearly puts these materials in the 
weak-coupling regime, with $U_{\rm real}< U_{\rm crit}$. Therefore, 
the stable orbital density wave in the chalcogen model with a finite 
but realistic Hubbard $U$ can be adiabatically connected to the 
ground state of the model without Coulomb interactions. We thus 
show that including a realistic value of inter-orbital Coulomb repulsion 
does {\it not} invalidate the paradigm of the 
`valence bond picture'~\cite{vonHippel1948, reitz1957,Matsui2014} 
[{\it i.e.} the `2-1-1' density wave of Fig.~\ref{charge-order}(a)] in the 
helical chalcogens:
The orbital density variations are already imposed by the combination 
of the helical chain structure and the anisotropic hopping amplitudes 
of the $p$ orbitals. The sole role of the relatively small 
inter-orbital Coulomb repulsion in the chalcogens is to explain the formation
of the helical chains themselves, starting from a hypothetical cubic crystal, 
as postulated in~\cite{silva2018}. 

It is only in the limit of unrealistically strong interactions with 
$U> U_{\rm crit}$ that we find a distinct orbital density wave, with 
only one type of orbital being unoccupied per chalcogen, i.e., the 
`2-2-0' density wave of Fig.~\ref{charge-order}(b). 
While the orbital density wave in the strong-coupling limit 
is triggered by the on-site Coulomb repulsion $U$, and can be easily 
understood in the fully localised limit of infinite $U$, the particular 
choice of the orbital which is occupied by a single hole is dictated by 
the kinetic energy.

\subsection{Relevance of the employed model and approximation}
Let us first comment on the role of the spin degree of freedom, 
neglected in this study. In some of the previous 
works~\cite{fukutome1984, shimoi1992, Deng2006, Deng2007} on the
subject it was postulated that the Hund's exchange could be the 
dominant mechanism which stabilises the double occupancy of one 
of the valence $p$ orbitals and partial occupancy of the other two 
orbitals on each chalcogen atom; in this way the `2-1-1' density 
wave should be easily supported in the chalocogens. While it is 
natural to expect that the Hund's exchange may support the `2-1-1' 
orbital density wave, this work, in combination with Ref. \cite{silva2018},
shows that, even without taking into account the electron's spin and
Hund's exchange, both the helical structure and the orbital density wave 
can be stabilised in elemental chalcogens. Although detailed further studies 
are needed here, the results shown here suggest that in the real materials 
the role of spin and Hund's exchange may also be secondary.  
In fact, to the best of our knowledge, there are no reports of any onset 
of spin density modulations in elemental chalcogens.
We remark that such modulations would indeed be expected if the spin 
degree of freedom mattered for the onset of the orbital density wave.

Next, let us discuss the validity of the Hartree approximation.
In this case it is worth pointing out that even in the case of the 
1D single-band (i.e., `standard') Hubbard model at half-filling, the Hartree
approximation leads to a partially correct result, 
especially in the weak-coupling limit~\cite{KhomskiiBook2014}.
Moreover, we expect that for the (anisotropic) orbital Hubbard model 
studied here, which lacks a continuous symmetry in the orbital sector, 
the Hartree approximation should work better, for the quantum fluctuations 
should then be somewhat suppressed.
Nevertheless, in order to check 
this presumtion, we performed preliminary DMRG calculations of the 
orbital Hubbard model, see Appendix~\ref{appb} for further details. 

Crucially, the obtained DMRG results unambiguously confirm that the `2-1-1' orbital density wave
is indeed stable in the weak-coupling limit---in particular, this density 
wave is the ground state well above the realistic value of $U=0.2t_\sigma$. 
On the other hand, the `2-2-0' orbital density wave, i.e. the density wave that is 
{\it not} observed in the chalcogens and that---according to the Hartree 
approximation---could become stable in the limit of unrealistically 
high Hubbard $U$ (see above), seems to be further destabilised in 
the DMRG simulations. In fact, according to the preliminary DMRG 
calculations, this density wave becomes stable only once a small 
crystal field, that has not been included in the model considered 
in this paper, but may nevertheless be present in the 3D chalcogens 
(see Appendix~\ref{appb}), is added. Note that, in order to unequivocally 
verify the stability of the `2-2-0' orbital density wave as well as to further 
corroborate the phase diagram of the orbital Hubbard model proposed 
here, further, extensive numerical studies are needed (due to the numerical 
complexity of the orbital Hubbard model these are beyond the scope of this work).

Finally, we note that the effects of the electron-phonon coupling are in general neglected
here and left for future studies. Nevertheless, we stress that the mere onset 
of the 1D helical chains in the chalcogens originates in a particular 
electron-phonon coupling, see~\cite{silva2018} and discussion in Sec.~\ref{sec:introcoul}.

\subsection{Final remarks}
The discussed here emergence of the different types of orbital order 
shows how the physics of the 1D $p$-orbital Hubbard model in 
a helical chain differs from that of the single-band Hubbard
model in one dimension which may host spin or charge density waves. 
In general terms, the reason for this is two-fold: First, the orbital 
systems are naturally prone to lattice distortions due to the strong 
coupling between orbitals and lattice. One should then consider the 
lattice distortions (such as the ones leading to helical chains in elemental
chalcogens) before deriving the physically-relevant, orbital Hubbard 
model. Second, unlike those in the single-band Hubbard model, the hopping 
amplitudes between distinct orbitals are generically strongly anisotropic, 
triggering spatial dependencies in observable quantities and phenomena.

\begin{acknowledgements}
We kindly acknowledge the support by the Narodowe Centrum Nauki 
(NCN, Poland) under Projects Nos. 
2016/22/E/ST3/00560 (A.K., C.E.A., and K.W.) and 
2016/23/B/ST3/00839 (A.K., A.M.O., and K.W.). 
We thank U. Nitzsche for technical assistance. 
C.E.A thanks S. Nishimoto for useful discussions.
\end{acknowledgements}

\begin{figure*}[t!]
  \flushleft{(a)} \hfill (b)\\
  \hskip .2cm
  \includegraphics[keepaspectratio=true,width=0.49\textwidth]{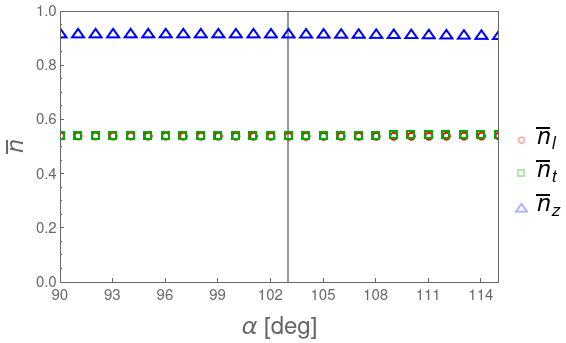}
  \includegraphics[keepaspectratio=true,width=0.49\textwidth]{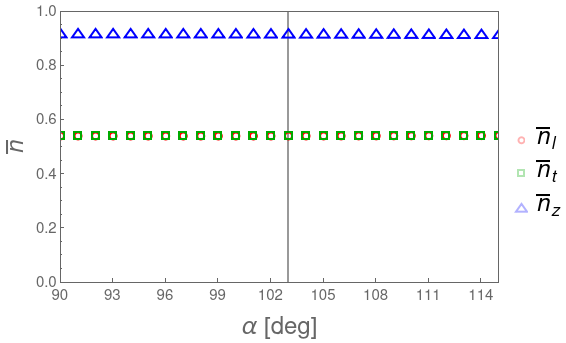}\\
  \flushleft{(c)} \hfill (d)\\
  \includegraphics[keepaspectratio=true,width=0.445\textwidth]{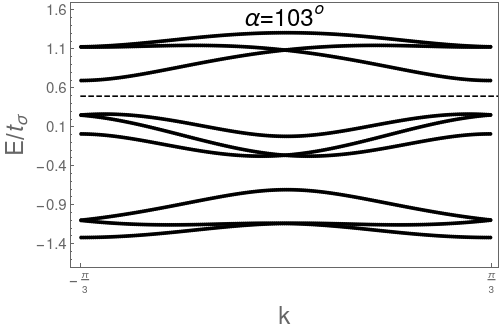}
\hskip .7cm
  \includegraphics[keepaspectratio=true,width=0.445\textwidth]{fig42l.png}
\caption{A comparison between the results with $U=0$, i.e., the 
non-interacting model, obtained using the full hopping matrix of 
Eq.~\eqref{full-hopping-matrix} [panels (a), (c)], and the linearized 
hopping matrix of Eq.~\eqref{hopping-matrix} [panels (b), (d)]. 
Top panels: the ground state orbital charge densities $\{\bar{n}_\mu\}$ in 
the local basis as functions of bond angle $\alpha$ (the bond angle 
$\alpha\approx 103^{\circ}$ observed in selenium and tellurium is 
marked with a vertical line). Bottom panels: the band structure for 
the bond angle $\alpha=103^{\circ}$ (dashed lines denote the Fermi 
energy). The hopping is $t_{\pi}=-t_{\sigma} / 3$ for all panels. 
The results are similar in both cases, with noticeable quantitative 
differences appearing only for bond angles considerably higher than the 
selenium/tellurium bond angle.}
  \label{linearized-vs-full}
\end{figure*}

\appendix

\section{The hopping matrix} 
\label{t_matrix}

In this Appendix, we derive the nearest neighbor hopping matrix 
${T}_{\mu,\nu}(i)$---here referred to as $\hat{T}(i)$. Since the chain 
has a helical symmetry, it suffices to consider a single bond to derive 
all the hopping elements in the chain or, phrasing it differently, in the 
local basis the hopping matrix $\hat{T}(i)$ is site-independent and we can
drop the site index: $\hat{T}(i)\equiv\hat{T}$ 
(as discussed in the main text). 

To describe a bond,
we consider two neighboring sites in the chain and label them $1$ and $2$. 
In order to find the local basis on site 2 one needs to take the local 
basis on site 1 and rotate it by $-{2\pi}/{3}$ around the helical axis, 
as shown in Fig.~\ref{geometry}. In the local basis, the helical axis is 
related to the local ${\mathbf z}$ axis via a rotation by $\beta$ around 
the local ${\mathbf t}$ axis [see Fig. \ref{geometry}(b)]. The angle 
$\beta$, in turn, can be written in terms of the bond angle $\alpha$ as:
\begin{align}
\cos\beta = \tan\left(\frac{\pi}{6}\right) 
\tan^{-1} \left(\frac{\alpha}{2}\right).
\end{align}
Consequently, the basis change from the local basis at site 1 to the 
local basis at site 2 is:
\begin{align}
\hat{R}_{1,2} = \hat{R}_t(\beta)\, \hat{R}_z\left(-\frac{2\pi}{3}\right)\, 
\hat{R}_t^{\dag}(\beta).
\end{align}
Finally, the vector pointing along the bond from site $1$ to site $2$ is 
rotated by $\left({\alpha}-{\pi}\right)/2$ around the local ${\mathbf z}$ axis
with respect to basis vector $\mathbf{l}$ of the local basis at site $1$. 

Since the $p$ orbitals transform like vectors under rotations, all of 
the above leads to the following expression for the hopping matrix
\begin{equation}
\hat{T}=\hat{R}_z\left(-\frac{\alpha}{2}+\frac{\pi}{2}\right)^{\dag}\, 
\hat{T}_0\,\hat{R}_z \left(-\frac{\alpha}{2}+\frac{\pi}{2}\right)\,
\hat{R}_{1,2}.
\end{equation}
Here $\hat{T}_0$ is the hopping matrix for the $p$ orbitals in 
a straight 1D chain, given by:
\begin{equation}
  \hat{T}_0 = \left(
  \begin{array}{ccc}
 t_{\sigma} & 0 & 0 \\
    0 & t_{\pi} & 0 \\
    0 & 0 & t_{\pi} \\
  \end{array}
  \right).
\end{equation}
The resulting hopping matrix for the helical chain is:
\begin{widetext}
\begin{equation} \label{full-hopping-matrix}
\left(
\begin{array}{ccc}
 \frac{1}{2} \left[t_{\sigma} (\sin(\epsilon )+1)+t_{\pi }\sin(\epsilon)\;\zeta^2\right] \;&\; 
\frac{1}{2}\,\zeta \left[t_{\sigma } (\sin(\epsilon )+1)-t_{\pi } \sin (\epsilon )\right] \;&\; 
t_{\pi }\, \sqrt{\frac{1}{-2 \sin (\epsilon )-2}+1}\,\zeta  \\
\frac{1}{2}\,\zeta \left[t_{\pi } \sin (\epsilon )-t_{\sigma } (\sin (\epsilon )+1)\right] \;&\;
\frac{1}{2} \left[t_{\sigma } (\sin (\epsilon )-1)-t_{\pi } \sin (\epsilon )\right] &\; 
t_{\pi }\,\sqrt{\frac{1}{-2 \sin (\epsilon )-2}+1} \\
 t_{\pi }\, \sqrt{\frac{1}{-2 \sin (\epsilon )-2}+1}\;\zeta \;&\; 
 -t_{\pi } \,\sqrt{\frac{1}{-2 \sin (\epsilon )-2}+1} \;&\; 
 -t_{\pi }\,\frac{1}{\csc (\epsilon )+1} \\
\end{array}
\right)\!. \notag
\end{equation}
\end{widetext}
Here $\zeta\equiv\cot\left(\frac{1}{4}(2 \epsilon +\pi )\right)$ and 
$\epsilon\equiv\alpha-\pi/2$. The linearized version of the hopping matrix 
is given in Eq.~\eqref{hopping-matrix} of the main text. For bond angles 
not much larger than $90^{\circ}$, the linearized model is sufficient 
to describe the band structure. The selenium/tellurium bond angle 
$\alpha=103^{\circ}$ lies comfortably within the range of applicability 
of the linearized model, as illustrated in Fig.~\ref{linearized-vs-full}.

\section{Chalcogenic Orbital Density Waves\\ 
         in Density Matrix Renormalization Group}
\label{appb}

\begin{figure}[h!]
    \centering
    {\hspace*{-8cm} (a)}
    
    \includegraphics[width=\columnwidth]{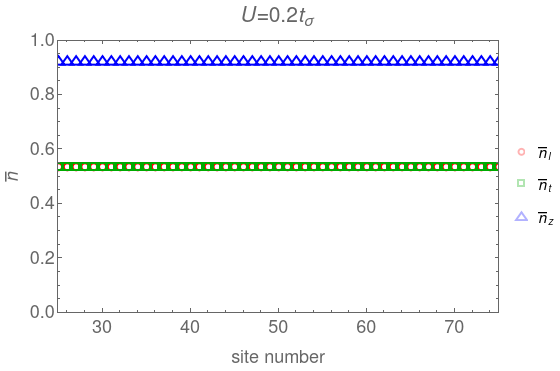}
    
    {\hspace*{-8cm} (b)}
    
    \includegraphics[width=\columnwidth]{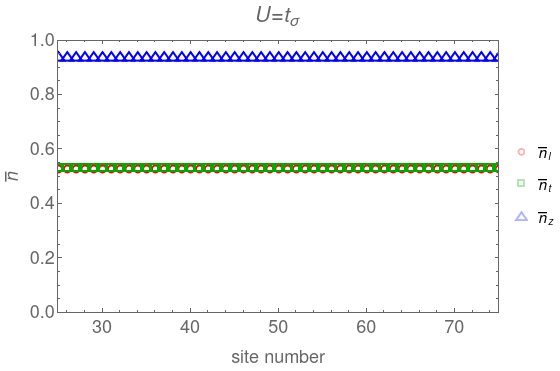}
\caption{Site-dependent ground state orbital occupations $\{\bar{n}_\mu\}$ 
in the local basis as obtained from DMRG calculations
of the Hubbard model (\ref{hamiltonian}) on
an $L=100$-site chain with: 
\mbox{$\text{(a)} \; U= 0.2 \; t_\sigma$,} \; \text{(b)} $U= t_\sigma$. 
Open boundary conditions are imposed and only densities on the 
`middle' 50 sites of the chain are shown.}
    \label{fig:dmrg-small-u}
\end{figure}

\begin{figure}[h!]
    \centering
    \includegraphics[width=\columnwidth]{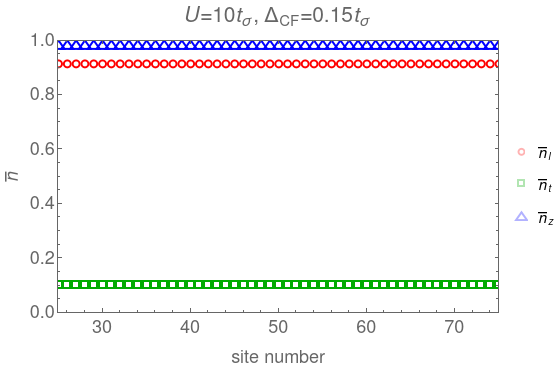}
\caption{Site-dependent ground state orbital occupations 
$\{\bar{n}_\mu\}$ in the local basis as obtained from DMRG calculations
of the Hubbard model (\ref{hamiltonian}) with 
$U=10 t_\sigma$ and with crystal field $\Delta_{\rm CF} =0.15 \; t_\sigma$
on an $L=100$-site chain (see text for further details). Open boundary conditions are imposed 
and only densities on the `middle' 50 sites of the chain are shown.}
    \label{fig:dmrg-large-u}
\end{figure}

In order to test the accuracy of the Hartree approximation (see main text), 
we performed preliminary DMRG calculations \cite{white1993} on systems of 
size $L\times 3$, $L$ being the number of sites in the chain and 3 being 
the number of orbitals per site. We use open boundary conditions. We fix 
$L=100$ in our calculations and keep up to $m=2000$ density-matrix eigenvalues 
in the renormalization procedure. This way we are able to obtain accurate 
results with an error $\delta/L=10^{-10}$. To suppress the edge effects, 
in what follows we plot the local orbital densities for all sites between 
site number $L/4$ and site number $3L/4$ in the chain.

\subsection{Weak Coupling}

In the weak coupling regime, extending to at least $U = t_\sigma$, 
we find that the DMRG results and the Hartree approximation results 
(see main text) are in perfect agreement, see~Fig~\ref{fig:dmrg-small-u}. 
This is due to the fact that the non-interacting system is in an 
orbitally-ordered phase protected by a finite energy gap, so that the 
(quantum) fluctuations in orbital densities are indeed negligible.

\subsection{Strong Coupling}

In the strong coupling regime we find that a small but finite 
symmetry-breaking crystal field $\Delta_{\rm CF}$ term must be 
included in the DMRG calculations to stabilise the `2-2-0' orbital density wave predicted 
by the Hartree approximation, see~Fig.~\ref{fig:dmrg-large-u}. 
Note that such a field, which raises the on-site energy of one of the 
orbitals in the bond-angle plane (e.g. the $p_t$ orbital as assumed here), 
may possibly arise in a model for a single chiral chain of the 3D 
chalcogen crystal once the influence of the neighboring chains 
(as arising from the electron tunneling or Coulomb interactions)
is 
taken into account. This is because, for bond angles greater than 
90$^\circ$, the presence of the neighboring chains needs to affect 
the on-site energies of the two orbitals lying in the bond angle plane 
differently [since one of them is oriented more along the chain ($p_l$)].

Without such a finite symmetry-breaking term the `2-2-0' orbital density wave predicted by
the Hartree approximation (see main text) is not recovered in our 
preliminary DMRG calculations for the Hubbard model with $U=10 \; t_{\sigma}$.
In fact, the calculation fails due to very long convergence time. 
To comment on this a little further, let us note that if one considers the extremely correlated regime ($U=20 \; t_\sigma$)
the calculation converges (unshown). In this case the result with zero crystal-field exhibits strong orbital density fluctuations in real space. We expect this to be---at least partially---the effect of the open edges. Importantly, in this extremely correlated regime we find a 
50-fold
increase in 
the CPU time for the calculation without the symmetry breaking field w.r.t. the calculation with the symmetry-breaking field included {\it or} the calculation for the weak coupling case ($U \leq t_{\sigma}$). 
This suggests the presence of competing interactions in the system 
and means that large-scale, state-of-the-art numerics 
are needed to establish the exact nature of the ground state without symmetry breaking field, possible order in the strong coupling limit or the dependence of the results on boundary conditions.


%

\end{document}